\documentstyle[twoside,fleqn,npb,epsfig]{article}
\newcommand{\AmS}{{\protect\the\textfont2
  A\kern-.1667em\lower.5ex\hbox{M}\kern-.125emS}}
\newcommand{\CLra}{{C_{Lq}^{N(1)}\over C_{Lg}^{N(1)}}}
\newcommand{\bCLra}{{\bar C_{Lq}^{N(1)}\over \bar C_{Lg}^{N(1)}}}
\newcommand{\CLrb}{{C_{Lq}^{N(2)} \over C_{Lg}^{N(1)}}}
\newcommand{\CLrc}{{C_{Lg}^{N(2)} \over C_{Lg}^{N(1)}}}
\newcommand{\Gqqa}{\gamma_{qq}^{(0)}}
\newcommand{\Gqga}{\gamma_{qg}^{(0)}}
\newcommand{\Ggqa}{\gamma_{gq}^{(0)}}
\newcommand{\Ggga}{\gamma_{gg}^{(0)}}
\newcommand{\Gqqb}{\gamma_{qq}^{(1)}}
\newcommand{\Gqgb}{\gamma_{qg}^{(1)}}
\newcommand{\Ggqb}{\gamma_{gq}^{(1)}}

\newcommand{\Cq}{C_{2q}^{N(1)}}
\newcommand{\Cg}{C_{2g}^{N(1)}}
\newcommand{\bb}{{\beta_1 \over \beta_0}}
\newcommand{\ZqqNT}{Z_{qq}^{NT}}
\newcommand{\ZqgNT}{Z_{qg}^{NT}}
\newcommand{\ZgqNT}{Z_{gq}^{NT}}
\newcommand{\ZggNT}{Z_{gg}^{NT}}

\title{Relations among Higher Order QCD corrections \thanks{Work supported in part by EU
contract FMRX-CT98-0194}}

\author{
J. Bl\"umlein\address{DESY-Zeuthen, Platanenallee 6, D-15738 Zeuthen, Germany.},
V. Ravindran$^{\rm a}$ 
and W.L. van Neerven\address {Instituut-Lorentz,
University of Leiden, P.O. Box 9506, 2300 HA Leiden, The Netherlands.}
}

\begin{document}

\begin{abstract}
We study the scheme transformation of next to leading order QCD 
corrections to various processes.  An interesting relation by Drell, 
Levy and Yan (DLY) among space like and time like processes is studied 
carefully in the next to leading order level.  We construct factorisation
scheme invariants and show that they are DLY--invariant.
\end{abstract}

\maketitle

\section{INTRODUCTION}

The Deep Inelastic Scattering (DIS) \cite{dis} of a lepton ($l$) on a hadron target 
(say proton $P$) is given by the process:
$l+P(p) ~\rightarrow~ l+ X$
where $p$ is the momentum of the target hadron.  The $X$ in the above process
denotes the final state hadrons which are summed over.  
The hadronic part which involves
the interaction of a virtual photon of momentum $q$ with virtuality $Q^2=-q^2$ 
with the hadron gives information
about the short distance structure of the hadron.  Usually, one studies this hadronic
part in terms of structure functions $F_i(Q^2,p\cdot q)$
($i=1,L$).  
%It is well known from the celebrated work of Bjorken \cite{bj1} that
%the functions scale in terms of a single variable,
From the parton model it follows that the functions only depend 
on the scaling variable
$x_B=Q^2/2p\cdot q$, when one considers the situation
where the scales $Q^2$ and $p \cdot q$ are very large \cite{bj1}.   
The parton model explains the scaling behavior in terms of what are known as
parton distribution functions.  
Due to the interaction between the partons in Quantum Chromodynamics(QCD), 
the scaling
is violated \cite{qcd} and hence these structure functions are no longer
just functions of $x_B$ alone, but dependent on the
scale $Q^2$ as well.  
The structure functions can be expressed in
terms of these quark and gluon distribution functions with appropriate
coefficients functions $C_{q,g}(z,Q^2 / M^2), M^2$ being the
factorisation scale.

The hadroproduction at $e^+e^-$ colliders
provide a wealth of information about how unobserved partons produced in the reaction
fragment into observed hadrons \cite{dly2}.  These cross sections are also
prametrized in terms of what are called fragmentation functions.
These functions in the QCD inspired parton model can be expressed
in terms of parton fragmentation functions convoluted with
parton level cross sections.  The corresponding scaling variable for this
process is defined as $x_E=2 p\cdot q/Q^2$, where $q$,~$p$ are the momenta
of photon of virtuality ($Q^2=q^2$) and the produced hadron respectively.  

\vspace{-0.3 cm}  
%\section{\bf Scheme transformation}
\section{SCHEME TRANSFORMATION}
The partonic cross sections computed in perturbative QCD suffer from
various divergences such as the infrared, ultraviolet and collinear singularities. 
For inclusive quantities all but the mass singularities cancel. The latter ones 
are absorbed into the parton distribution functions  
at a factorisation scale $M^2$.
In practice, one first separates
the singular part of the partonic cross sections into a 
process independent function, called
transition function $\hat \Gamma(1/\epsilon,\alpha_s(R^2),M^2/\mu^2,M^2/R^2,scheme)$,
which contains only the mass singularities.  
This object is then convoluted with the bare parton distributions
and the resulting ones are called renormalized parton distributions.
Hence, the perturbatively calculable coefficient functions computed
in QCD suffer from scheme dependency.  
The scheme dependency appearing in the coefficient
functions and the parton distribution functions are expected to cancel
since the convolution of them, 
which is the structure function, is a physical
observable and is thus scheme independent \cite{scheme}.
The above discussion holds for
time like processes as well.

Consider any two physical observables denoted by $F_A^N(Q^2)$ and $F_B^N(Q^2)$:
\begin{equation}
F_I^N(Q^2)=\int_0^1 dx x^{N-1} F_I(x,Q^2) \quad \quad I=A,B
\end{equation}
For simplicity we 
consider the singlet case only.
The general case is dealt with in Ref.~\cite{new}
\begin{eqnarray}
F_I^N(Q^2)&=&f_q^{(S)N}(Q^2) C_I^{(S)N}(Q^2)
\nonumber \\
&&+ g^N(Q^2) C_I^{(S)N}(Q^2),
\end{eqnarray}
with
\begin{equation}
C^{(S)N}_{Ii}(Q^2)=\int_0^1 dz z^{N-1} C_{Ii}^{(S)}(z,Q^2) ,
\end{equation}
where $i=q,g$.
Using the fact that $f_q^{(S)N}(Q^2)$ and $g^N(Q^2)$ satisfy renormalization group
equations (RGE), 
defining
\begin{equation}
t=-{2 \over \beta_0} \log\left({a_s(Q^2) \over a_s(Q_0^2)}\right),
\end{equation}
where $a_s(Q^2)=\alpha_s(Q^2)/4 \pi$,
expanding the anomalous dimensions and coefficient functions
in terms of the strong coupling constant
one finds that 
\begin{equation}
%\begin{displaymath}
{\partial \over \partial t}\left( \begin{array}{c}
F^N_A\\
{F^N_B}
\end{array} \right)= -{1 \over 4}
\left( \begin{array} {cc}
\Gamma^N_{AA} & \Gamma^N_{AB}\\
\Gamma^N_{BA} & \Gamma^N_{BB} 
\end{array} \right)
\left( \begin{array}{c}
F^N_A\\
{F^N_B}
\end{array} \right).
%\end{displaymath}
\end{equation}
The matrix entries $\Gamma^N_{IJ}$ depend on the anomalous dimensions
and the coefficient functions.
Though the anomalous dimensions
and coefficient functions depend on the scheme in which they are computed,
the entries $\Gamma^N_{IJ}$ constructed out of them are scheme independent,
because the structure functions $F_I^N(Q^2)$ are physical observables which do not depend on the
factorisation scheme.  This property is studied in the forthcoming sections.

In general, the functions
$\Gamma^N_{IJ}$ have the following expansion in terms of coupling constant:
\begin{eqnarray}
\Gamma_{ij} = \sum_{l=0}^{\infty} a_s^l(Q^2) \Gamma_{ij}^{(l)} ,
\end{eqnarray}
where we have suppressed the $N$ dependence.
Consider the choice where 
\begin{equation}
F^N_A(Q^2)=F^{N(S)}_2(Q^2),F^N_B(Q^2)=\displaystyle {F^N_L(Q^2) 
\over a_s(Q^2) C_{Lg}^{(1)} }
\end{equation}
For convenience, $F^N_B(Q^2)$ is normalized by a factor $a_s(Q^2) C_{Lg}^{(1)}$ which is
factorisation scheme independent. 
Since the leading order terms are made out of leading order anomalous dimensions
and coefficient functions, they are 
scheme invariants.  To next to leading order in $a_s(Q^2)$, the entries are 
lengthy, so we present
here $\Gamma_{22}^{(1)}$ only (see Ref.~\cite {new} for the
other entries): 
\begin{eqnarray}
&&\Gamma_{22}^{(1)}=\Gqqb-\bb \Gqqa-\CLra \left(\Gqgb
\right.
\nonumber \\
&&\left. -\bb \Gqga\right) 
+\CLra \Cg \Gqqa \nonumber \\
&&-\left[\CLrb+\left(\CLra\right)^2 \Cg
\right.
\nonumber \\
&&\left. -\CLra \CLrc\right] \Gqga
+\Cg \Ggqa \nonumber \\
&& -\CLra \Cg \Ggga + 2 \beta_0 \Bigl(\Cq
\nonumber \\
&& -\CLra \Cg\Bigr),
\end{eqnarray}

The form of the time like 
$\Gamma_{IJ}$'s is same as that in the space like case but with
obvious changes such as $\gamma_{ij}$ and $C^N_{Ij}$~ are replaced by
the corresponding time like ones.  One verifies 
that the $\Gamma_{IJ}$'s~ are invariant under factorisation scheme transformations
and hence they are physical observables.  
It is worth emphasising the
fact that the scheme dependency coming from the two loop anomalous
dimensions cancels exactly the scheme dependency
coming from the ${\cal O}(a^2_s(Q^2))$ coefficient functions.

%\section{\bf Drell-Levy-Yan relations}
\section{DRELL-LEVY-YAN RELATIONS}
In this section we study in detail, an interesting relation between
deep inelastic lepton hadron scattering and $e^+e^-$ 
annihilation to a hadron plus
anything else, proposed by Drell, Levy
and Yan \cite{dly2}.  According to their work,  
%are same as those defined in the beginning of the paper.
if the Bjorken limit exists for both DI scattering
and DI annihilation,
then the scaling structure functions satisfy the following relation:
\begin{equation}
F_i^T(x_E)=F_i^S(1/x_B)\quad \quad  i=1(T),2. 
\end{equation}
In other words,
$F_i^T(x)$ are the continuations of the corresponding
functions $F_i^S(x)$ from $x <1$ to $x >1$.  This is true only when
the continuation is smooth, i.e. there are no singularities, for example
at $x=1$, or others.  This relation is called DLY relation in the literature.
 
In this section, we study this property in more detail \cite{new,brv}.  
As we know, the splitting functions and coefficient functions are not
physical due to the scheme choice to renormalise mass singularities.
Hence, the naive continuation rule is violated in general.  
It was demonstrated in the paper by Curci, Furmanski and Petronzio \cite{cfp} that
by appropriately modifying the continuation rule in the $\overline {\rm MS}$
scheme, one can show that the time like splitting functions are related
to space like counter parts.  Since the modification of the continuation
rule is related to the scheme one adopts, it simply amounts to
finding finite renormalisation factors for these quantities.  It was shown that the finite
renormalisation factors can be constructed from the 
$\epsilon$--dependent 
part of the splitting function when computed in dimensional regularisation \cite{sv}.   
In addition to this, care should be taken when dealing with quarks and gluonic states
which was not the case yet in the DLY work.   It amounts to multiplying by $-1$ for continued 
space like $P_{qq},P_{gg}$, $C_F/(2 n_f T_f)$ for $P_{qg}$, 
and $2 n_f T_f/C_F$
for $P_{gq}$.  This is independent of the scheme because it results from the
crossing of the particles between $in$ and $out$ states.
Keeping this in mind and using the known splitting functions \cite{cfp,fp} one finds that
\begin{eqnarray}
P_{ij}&=&\sum_{\{k,l\}=q,g}Z^T_{ik} \otimes \bar P_{kl}\otimes (Z^{T-1})_{lj}
\nonumber \\
&&            -2 \beta(a_s)\sum_{l=q,g}  Z^T_{il}\otimes {d \over d a_s} (Z^{T-1})_{lj},
\label{eq9}
\end{eqnarray}
where the quantities with a bar on the top denote that they are continued from $z \rightarrow 1/z$
with appropriate factors in front.  
The relations given above remain true for the polarized splitting 
functions 
\cite{mvv,sv} as well.
The renormalisation factors are given by
\begin{eqnarray}
Z^T_{ij}&=&P_{ji}^{(0)} \Big(\log(z) + a_{ji}\Big).
\end{eqnarray}
The terms $a_{ij}$ depend on whether polarized or unpolarized splitting functions
are considered.
In the case of unpolarized splitting functions, one finds that
\begin{equation}
a_{qq}=a_{gg}=0,\quad  a_{qg}=-1/2,\quad a_{gq}=1/2.
\end{equation}
For the polarized case, one obtains
\begin{equation}
a_{ij}=0.
\end{equation}
The $\log$ in the renormalisation factors originates from the kinematics
and the factor $\pm 1/2$ come from the polarisation averaging of the gluons.

At this point it is worth comparing with the work of Gribov and Lipatov \cite{gribov} (GL)
in which    
$P_{qq}^{(S)}(z)= P_{qq}^{(T)}(z)$ 
is claimed.
This relation is preserved at the leading order in $a_s$ but broken
at higher orders.  Using the method of \cite{cfp}, the breaking
terms can be identified with those coming from the ladder diagrams beyond
leading order.  Ref. \cite{cfp} shows that the GL relation is broken
even for the physical quantities unlike the DLY-relation by considering scheme
invariant combination of non--singlet 
structure functions.  In this section
we will substantiate their result by looking at the singlet scheme invariant
physical quantities.

Now, let us study how space like and time like coefficient functions
are related.  The coefficient functions are nothing but
the parton level cross sections renormalized by mass factorisation.
Hence it is expected to violate the DLY relation due to the
scheme in which they are computed.  
The leading order longitudinal coefficient functions are identically zero.
The next to leading order longitudinal ones 
do not get any correction.  
The reason for this is that the unrenormalised longitudinal coefficient functions
do not contain any mass singularities unlike the transverse coefficient functions,
hence there is no left over finite piece which could arise from the $z^\epsilon$
terms
or the $n-$diminsional polarisation average.  
At NNLO level, the longitudinal coefficient functions alone do not satisfy the DLY relation
anywhere.
We follow the results given in \cite{sanchez,zv2,zijlstra} for the space like
and \cite{rv2,rijken} for the time like case.
It turns out that they are related by the $Z$ factors in a 
non--linear way
as given below.  
\begin{eqnarray}
&&\!\!\!\!C_{Lq}^{(2)T}(z)\!-\!{z \over 2}  C_{Lq}^{(2)S}\left({1 \over z}\right)
=
\!Z^T_{qq}\otimes {z \over 2} C_{Lq}^{(1)S}\left({1 \over z}\right)
\nonumber\\
&&~~~~~~~+Z^T_{gq}\otimes {C_F \over 2 n_f T_f} {-z \over 2}
C_{Lg}^{(1)S}\left({1 \over z}\right)
\nonumber \\
&&{1 \over 2} \left[C_{Lg}^{(2)T}(z)\!-\!{C_F \over 2 n_f T_f}{z \over 2} 
                                   2 C_{Lg}^{(2)S}\left({1 \over z}\right)\right]
\nonumber \\
&&~~~~~~~~~~\!=\!
Z^T_{qg}\otimes {z \over 2} C_{Lq}^{(1)S}\left({1 \over z}\right)
\nonumber \\
&&~~~~~~~+Z^T_{gg}\otimes {C_F \over 2 n_f T_f} {-z \over 2}  
C_{Lg}^{(1)S}\left({1 \over z}\right).
\label{eq13}
\end{eqnarray}
The right hand side of the above equation contains the various convolutions
of $Z$ factors with the continued NLO longitudinal space like coefficient functions.  

Let us define
the difference between time like quantities ($\Gamma_{ij}^T$)and
continued space like quantities($\Gamma_{ij}^S$) as 
$\delta \Gamma_{ij}=\Gamma_{ij}^T - \bar \Gamma_{ij}^S$.
Using the $N$th moment of Eq.~(\ref{eq13}) , we get
\begin{eqnarray}
 \delta \Gamma_{22}^{(1)}&=& \delta \Gqqb - 2 \beta_0 \ZqqNT - \bar \Ggqa  \ZqgNT
\nonumber \\
&&+\bar \Gqga \ZgqNT
+\bCLra \Big( -\delta \Ggqb 
\nonumber \\
&&+2 \beta_0 \ZqgNT 
-\ZqgNT \bar \Gqqa + \ZqqNT \bar \Gqga 
\nonumber \\
&&  -\ZggNT \bar \Gqga +\ZqgNT \bar \Ggga \Big).
\end{eqnarray}
Eq.~(\ref{eq9}) implies that
$\delta \Gamma_{22}^{(1)} = 0$.  The same is true for other entries of the
$\Gamma$ matrix.
It is clear from
the above exercise that for time like physical anomalous dimensions $\Gamma_{ij}^T$, one
can directly use the space like physical anomalous dimension with the
appropriate changes such as $z \rightarrow 1/z$ and the corresponding
changes in the overall colour factors $without$ using any $Z$ factors.  
Also, the GL relation cannot be thought of as scheme transformation, hence
the physical anomalous dimensions are not preserved in this case.

\end{document}